\documentclass[pre,a4paper,superscriptaddress,showpacs]{revtex4}
\usepackage{graphicx}
\begin{document}

\newcommand{\EQ}{Eq.~}
\newcommand{\EQS}{Eqs.~}
\newcommand{\FIG}{Fig.~}
\newcommand{\FIGS}{Figs.~}
\newcommand{\TAB}{Tab.~}
\newcommand{\SEC}{Sec.~}
\newcommand{\SECS}{Secs.~}

\title{Transmission of severe acute respiratory syndrome
in dynamical small-world networks}
\author{Naoki Masuda}
\author{Norio Konno}
\affiliation{Faculty of Engineering,
Yokohama National University,
79-5, Tokiwadai, Hodogaya, Yokohama, 240-8501 Japan}
%
\author{Kazuyuki Aihara}
\affiliation{Department of Complexity Science and Engineering,
Graduate School of Frontier Sciences,
University of Tokyo, 7-3-1 Hongo Bunkyo-ku Tokyo 113-8656 Japan}
\affiliation{ERATO Aihara Complexity Modelling Project,
Japan Science and Technology Agency, Tokyo, Japan}
\date{\today}

\begin{abstract}
The outbreak of severe acute respiratory syndrome (SARS) is still
threatening the world because of a possible resurgence.  In the
current situation that effective medical treatments such as antiviral
drugs are not discovered yet, dynamical features of the epidemics
should be clarified for establishing strategies for tracing,
quarantine, isolation, and regulating social behavior of the public at
appropriate costs. Here we propose a network model for SARS epidemics
and discuss why superspreaders emerged and why SARS spread especially
in hospitals, which were key factors of the recent outbreak. We
suggest that superspreaders are biologically contagious patients, and
they may amplify the spreads by going to potentially contagious places
such as hospitals. To avoid mass transmission in hospitals, it may be
a good measure to treat suspected cases without hospitalizing them.
Finally, we indicate that SARS probably propagates in small-world networks
associated with human contacts and that the biological
nature of individuals and social group properties are factors
more important
than the heterogeneous rates of social contacts among individuals. This is
in marked contrast with epidemics of sexually transmitted diseases
or computer viruses to which scale-free network models often
apply.
\end{abstract}


\maketitle

\section{Introduction}\label{sec:introduction}

The first case of the recent outbreak of severe acute respiratory
syndrome (SARS) is estimated to have started in the Guandong province
of the People's Republic of China in November of 2002.  After
that, SARS spread to many countries, causing a number of infectious
cases. In spite of worldwide research efforts, the biological
mechanism of the SARS infection is not yet fully clarified, which mars
developments of antiviral drugs or other means of conclusive
medication.  Under this condition, an effective way was to track
everybody suspected to be involved in the spreads and quarantine them,
which is the same as a century ago. However, more effective strategies
in terms of safety and cost could be established with the knowledge of
dynamical mechanisms of the outbreak including the effects of so called
superspreaders (SS's) and spreads in hospitals.  Along this line,
epidemiological models that explain the actual and potential
transmission patterns can be helpful for suppressing the spreads.  For
example, dynamical compartmental models for fully mixed population
\cite{Lipsitch} and for geographical subpopulations in Hong Kong
\cite{Riley} have been proposed and fitted to the real data, and they
are successful in explaining the real data and determining the basic
reproductive number \cite{May_NAT_ROYAL}.  However, the models contain
many compartments and many parameters whose values are determined
manually, which may obscure relative contributions of the
factors. Here we rather propose a simplified spatial model to indicate
how interplay between network structure and individual factors affects
the epidemics.

A prominent feature in the SARS epidemics is the dominant influence of
SS's \cite{Lipsitch,Riley,Singapore}.  According to
the US 
Centers for Disease Control and Prevention (CDC), a patient is
defined to be a SS if he or she has infected more than 10 people. The SARS
epidemics are
special in that a majority of cases originated from just a small
number of SS's. On the other hand, nonsuperspreading patients, which
by far outnumber SS's, explain only a small portion of the infection
events. In Singapore, just 5 SS's
have infected 80\% of about 200 patients, whereas about 80\% of
the patients have infected nobody
\cite{Singapore,Vogel_NAT,Abdullah}.
Also in Hong Kong, one patient caused more than 100 successive cases
\cite{Riley,Abdullah}.
Similar key persons are identified in other parts of the world as well.
Also epidemics of Ebola, measles,
and tuberculosis often accompany SS's \cite{Singapore}. It is believed
that SS's are caused both by biological reasons such as genetic
tendencies, health conditions, and strength of the virus and by
social reasons such as the manner of social contacts and global
structure of social interaction. It agrees with general understanding
that epidemics depend on the personal factors and the structure of social
networks \cite{Keeling,Newman_PRE02}.  Although previous dynamical
models consider SS's to be exceptional \cite{Riley} or do not model them
explicitly \cite{Lipsitch}, we incorporate them as a key factor for
the spreading.

Another feature of SARS is rapid spreading in hospitals, which
played a pivotal role in, at least, local outbreaks, sometimes
accounting for more than half the total regional cases.  The
embarrassing fact that hospitals are actually amplifying diseases
\cite{Riley,Singapore} should be provided with convincing mechanisms
so that we can reduce the risk of 
spreads in hospitals and relieve the public of
anxieties. To this end again, we will examine the combined effects
of SS's and the network structure.

Here we construct a dynamical model for SARS spreads, which is simpler
than the previous models \cite{Lipsitch,Riley} but takes into account
SS's and the spatial structure represented by the small-world properties
\cite{Watts_Wattsbook}. We then propose possible means for preventing
SARS spreads in the absence of vaccination. The simulated SARS
epidemics are also compared with the epidemics of sexually transmitted
diseases (STD's) and computer
viruses whose mechanism owes much to scale-free properties of the
underlying networks \cite{Newman_PRE02,Barabasi_SCI,Liljeros,Pastor}.

\section{Model and General Theory}

Our model is composed of $n$ persons located on vertices of a graph.
A pair of individuals connected by an undirected edge directly
interact and possibly transmit SARS.  We simply assume three types of
individuals: namely, the susceptible, the infected but non-SS's, and
the SS's. Here a SS, probably with strong and/or a large amount of
viruses, has a strong tendency to infect the susceptible, even without
frequent social contacts. The dynamics is the contact
process with three states \cite{Pastor,Durrettbook,Schinazi_MB01}. A
susceptible can be infected by an adjacent patient (a SS or
an infected non-SS)
at certain rates. A patient returns to the susceptible state
at rate 1, mimicking the recovery from SARS or its death followed by
the local emergence of a new healthy person. The infected non-SS's and SS's are
modeled with different rates of infection
\cite{May_NAT_ROYAL,Newman_PRE02,Schinazi_MB01}.  An infected turns an
adjacent susceptible into infected non-SS or SS at rate $\lambda_I (1-p)$ or
$\lambda_I p$, respectively, where $p$ parametrizes the number of SS's
divided by the number of patients.  Similarly, a SS infects an
adjacent susceptible into infected non-SS or SS at a rate $\lambda_{SS} (1-p)$
or $\lambda_{SS} p$, respectively \cite{Schinazi_MB01}. The infected non-SS's
and SS's do not have direct interactions even if they are next to each
other. However, they interact indirectly owing to the cross-talk rates
$\lambda_I p$ and $\lambda_{SS} (1-p)$. These infection events as well
as death events at rate 1 happen independently for all the sites.  The
parameter values depend on the definition of a SS, the network
structure, and the time scales.  With the supposition of total mixing
of the individuals and the definition of a SS by CDC, the data of the
outbreak in Singapore \cite{Singapore} provide a rough estimate of
$p=0.03$.  As a rough estimation, we
set $\lambda_{SS}/\lambda_I = 20$ based on the descriptions on 
a small number of
superspreaders identified in Singapore \cite{Singapore} and
Hong Kong \cite{Riley,Abdullah}. 
To our knowledge, larger data about the number
of cases caused by each patient or about the detailed chains of
transmissions are not available in other regions.  A relevant condition
that seemingly holds in the current outbreak is $\lambda_I < 1 <
\lambda_{SS}$, where $\lambda_I$ and $\lambda_{SS}$ are multiplied by
the number of neighbors for a moment.  In this situation, the
mean-field theory predicts the existence of a threshold for $p$ above
which the disease spreads widely \cite{Schinazi_MB01}. The recent
outbreak may have led to a suprathreshold regime even with small
$p$ because $\lambda_{SS}$ is presumably huge. The model studies using
real data suggest that the threshold has been crossed from the
above by the control efforts \cite{Lipsitch,Riley}.

Next, we introduce the local network structure. At a given time, the
whole population is typically divided into groups within which
relatively frequent social contacts are expected. A group
represents, for example, hospital, school, family,
market, train, and office, and it
is characterized by clustering properties
\cite{Watts_Wattsbook,Amaral_PNAS} and dense coupling.
We prepare $g$ groups, each containing $n_g =
n/g$ individuals. The $i$th individual ($1\le i\le n$) is connected to
randomly chosen $k_i$ ($0\le k_i \le n_g-1$) individuals within the
group.  The rate of transmission is proportional to the vertex degree
$k_i$ in the early stage of epidemics \cite{May_NAT_ROYAL,Pastor}.
Apart from the effects of $k_i$, $\lambda_I$, and
$\lambda_{SS}$, some social groups are more prone to transmit SARS
than others. This group dependence originates in, for example,
ventilation, sanitary levels, and the duration of grouping
\cite{Lipsitch,Riley,Vogel_NAT}. The effect
is represented by a multiplicative factor $T_j$ for the $j$th
group ($1\le j\le g$). Then the effective intragroup 
infection strength is calculated as $\left<k_i\right>_j T_j$, where
$\left<\cdots\right>_j$ is the average over $i$ in the $j$th group.  Presumably,
social groups such as hospitals, congested trains, airplanes, and
poorly ventilated residences have large $\left<k_i\right>_j T_j$.  For
example, hospitals may have large $\left<k_i\right>_j T_j$ because of
a high population density yielding large $\left<k_i\right>_j$ and
the fact that the susceptible hospitalized for other diseases may
be generally weak against infectious 
diseases including SARS.  The
influence of trains due to congestion and closedness of the air for
long time is a potential
source of outbreaks in the regions where people habitually commutate
by congested public transportations, like Japan.  In contrast,
$\left<k_i\right>_j T_j$ may be low for groups formed in open spaces.
However, we note
that SARS can also break out in low-risk groups if $\lambda_{SS}$ is
sufficiently large.  For simplicity, we assume that $g_0$ out of $g$
groups have $T_j=T_h$ that is larger than $T_j=T_l$ taken by the other
$g-g_0$ groups.

Although many models ignore the spatial structure of the population and rely
on meanfield descriptions \cite{Lipsitch,May_NAT_ROYAL}, spatial
aspects should be incorporated for understanding the real dynamics of
epidemics \cite{Riley,Keeling,Newman_PRE02,Lloyd_SCI}.
Mainstream from this standpoint are methods of percolation and
the contact process on regular lattices
\cite{Schinazi_MB01,Durrettbook,Liggettbook_Schinazibook}. However,
$d$-dimensional 
lattices have characteristic path length $L$ --- that is, the mean
distance between a pair of vertices --- proportional to
$n^{1/d}$. In social networks, $L$ is approximately
 proportional to $\log n$ as in random
graphs \cite{Watts_Wattsbook}. To cope with this observation, we introduce
random recombination of $n$ individuals into $g$ new groups. In
reality, one belongs to many groups that dynamically break and reform
more or less randomly by way of social activities
\cite{Keeling,Watts_SCI}. For example, one may commute to one's workplace
and return home everyday, possibly by changing trains, which serve
as temporary social groups as well.  After time $t_0$, we randomly
shuffle all the vertices and reorganize them into $g$ groups and wire the
vertices within each group in the same manner as before. Then the
epidemic dynamics is run for another $t_0$ before next shuffling
occurs. For simplicity, just two independent groupings are assumed to
alternate, as schematically shown in \FIG\ref{fig:network}. However,
the results are easily extended to the case of longer chains of group
reformation.  Owing to the shuffling, individuals initially belonging to
different groups can interact in the long run.

We denote $x_{\alpha,I}$ and $x_{\alpha,SS}$ the number of the
infected non-SS's and that of the SS's summed over the groups with $T_j =
T_{\alpha}$ ($\alpha=h$, $l$).  In the early stages of epidemics, the
dynamics between two switching events is given by the meanfield
description as follows:
\begin{equation}
\begin{tiny}
\frac{d}{dt} \left( \begin{array}{c}
x_{h,SS}\\ x_{h,I}\\ x_{l,SS}\\ x_{l,I}
\end{array} \right) =
\left( \begin{array}{cccc}
\lambda_{SS} p \left<k_i\right>_h T_h -1 & \lambda_I p \left<k_i\right>_h T_h & 0 & 0\\
\lambda_{SS} (1-p) \left<k_i\right>_h T_h & \lambda_I (1-p) \left<k_i\right>_h T_h -1 & 0 & 0\\
0 & 0 & \lambda_{SS} p \left<k_i\right>_l T_l - 1 & \lambda_I p \left<k_i\right>_l T_l\\
0 & 0 & \lambda_{SS} (1-p) \left<k_i\right>_l T_l & \lambda_I (1-p) \left<k_i\right>_l T_l - 1
\end{array} \right)
\left( \begin{array}{c}
x_{h,SS}\\ x_{h,I}\\ x_{l,SS}\\ x_{l,I}
\end{array} \right),
\end{tiny}
\label{eq:dynamics}
\end{equation}
where $\left<\cdots\right>_{\alpha}$ denotes averaging over the groups with
$T_j = T_{\alpha}$. The random shuffling is expressed by
multiplication of the following matrix from the left:
\begin{equation}
\left( \begin{array}{cccc}
\frac{g_0}{g}+\sigma & 0 & \frac{g_0}{g}+\sigma & 0\\
0 & \frac{g_0}{g}+\sigma & 0 & \frac{g_0}{g}+\sigma\\
\frac{g-g_0}{g}-\sigma & 0 & \frac{g-g_0}{g}-\sigma & 0\\
0 & \frac{g-g_0}{g}-\sigma & 0 & \frac{g-g_0}{g}-\sigma
\end{array} \right),
\end{equation}
where $\sigma$ is the possible correlation factor specifying the
tendency for patients to join groups with $\left<k_i\right>_j T_j =
\left<k_i\right>_h T_h$.  Purely random mixing yields $\sigma=0$. The
map for the one-round dynamics comprising the contact process for time
$t_0$ followed by switching has eigenvalues 0, 0, ${\rm e}^{-t_0}\cong
1-t_0$, and
\begin{eqnarray*} 
&& (\frac{g_0}{g}+\sigma) {\rm
e}^{(-1+T_h\left<k_i\right>_h(\lambda_I(1-p)+\lambda_{SS}p))t_0} +
(\frac{g-g_0}{g}-\sigma) {\rm
e}^{(-1+T_l\left<k_i\right>_l(\lambda_I(1-p)+\lambda_{SS}p))t_0}\\
&\cong&  1+ \left\{ \left[ \left(\frac{g_0}{g}+\sigma\right)
T_h\left<k_i\right>_h
+ \left(\frac{g-g_0}{g}-\sigma\right)
T_l\left<k_i\right>_l\right] \left[\lambda_I
(1-p) + \lambda_{SS} p\right] -1 \right\} t_0
\end{eqnarray*}
for $t_0$ small with
respect to the system time $t$ introduced in \EQ(\ref{eq:dynamics}). An
important indicator of outbreaks is the basic reproductive number
$R_0$ defined as the mean number of secondary infections produced by a
single patient in a susceptible population
\cite{Lipsitch,Riley,May_NAT_ROYAL,Keeling,Newman_PRE02,May_PRE}.  If
$R_0$ exceeds unity, the disease spreads on average in mixed
populations such as the local groups in \FIG\ref{fig:network}.  Since
$R_0$ equals the largest eigenvalue, what matters is whether
\begin{equation}
\left[(\frac{g_0}{g}+\sigma)T_h\left<k_i\right>_h +
(\frac{g-g_0}{g}-\sigma)T_l\left<k_i\right>_l\right] \left[\lambda_I
(1-p) + \lambda_{SS} p\right]\nonumber
\end{equation}
is greater than 1. As a result,
multiple kinds of heterogeneities \cite{May_NAT_ROYAL} --- namely, the
factors associated with individual patients and those specific to the
groups --- interact and determine the tendency to spread. Generally
speaking, a positive $\sigma$ raises $R_0$.  Even if both factors are
subthreshold in the absence of $\sigma$, that is
\begin{equation}
\left(\frac{g_0}{g}
T_h \frac{\left<k_i\right>_h}{\left<k_i\right>} + \frac{g-g_0}{g}
T_l\frac{\left<k_i\right>_l}{\left<k_i\right>} \right) < 1\nonumber
\end{equation}
and
$\left[\lambda_I (1-p) + \lambda_{SS} p\right]\left<k_i\right> < 1$, a
positive $\sigma$ can make the whole dynamics suprathreshold.  In
actual SARS spreads in hospitals; $\sigma>0$ seems to have held;
compared with healthy people, the SARS patients and the suspected are
obviously more likely to go to hospital where $T_j$ and
$\left<k_i\right>_j$ are supposedly high. Currently, we do not have
control over infection rates of individuals, particularly
$\lambda_{SS}$ \cite{Riley}.  However, the threat of spreads may be
decreased if their behavior is altered so that they avoid risky
places. It is recommended that they be seen by doctor at home or some
isolated sites. The strategies applied in many countries such as
introducing more separated hospital rooms, making doctors and nurses
work in a single ward \cite{Meyers}, and ordering the public to stay
home also decrease $k_i$ and $\sigma$ \cite{Riley}.

\section{Simulation Results}

We next examine effects of network structure by numerical
simulations. To focus on
topological factors, we simply set $T_h =
T_l = 1$ and $k_i = k = n_g-1$ ($1\le i\le n$).  The group size $n_g$,
which is typically somewhat 
smaller than 100 \cite{Watts_SCI}, is chosen to be
$81=9^2$ for technical reasons,
although the value really relevant to the SARS epidemics is not
known \cite{Lipsitch}. With $g=100$, $n=g n_g = 90^2$, and
$t_0=0.5$, the chains of
infection after the total run 
time $\overline{t}=1.0$, from the viewpoint of two different groupings as in 
\FIG\ref{fig:network}, are shown in
\FIGS\ref{fig:map}(a) and \ref{fig:map}(b).
They more or less reproduce the transmission
pattern of SARS in Singapore \cite{Singapore}, including the rapid
spreads mediated by small $L$ and the massive influence of
SS's (solid lines). The
transmission naturally spreads over time, as shown in
\FIG\ref{fig:map}(c) corresponding to
$\overline{t}= 2.0$.
By comparing \FIG\ref{fig:map}(c) with
\FIG\ref{fig:map}(d),
which shows the results for $\overline{t}=2.0$ and $t_0=1.0$, 
we find that local transmission develops if
the time spent with a fixed group configuration is relatively longer.

More quantitatively, \FIG\ref{fig:trans}(a) shows, for
$\overline{t}=2.0$ and $t_0=0.5$,
the distributions of $a_i$, which is the number of people to whom the
$i$th patient has directly infected. The patients with large $a_i$ are
mostly SS's. Small $a_i$ is chiefly covered by other patients, and the
distribution decays exponentially in $a_i$ within this range.
The homogeneous vertex degree and the Poiss\'{o}n
property of the processes caused 
the exponential tail, which is preserved in
small-world-type networks like ours
and random graphs \cite{Watts_Wattsbook} where the vertex
degrees obey narrow distributions.

\section{Discussion}\label{sec:discussion}

\subsection{Comparison with regular lattices}

A time course of chains of infection in a two-dimensional square
lattice are shown in \FIGS\ref{fig:map}(e), \ref{fig:map}(f), and
\ref{fig:map}(g), with $n$, $g$, and $k_i$, and the duration of the
run the same as before. We assume the periodic boundary conditions,
and $k_i=80$ neighbors of a vertex ($x$,$y$) ($1\le x,y\le 90$) are
defined to be the vertices included in the square with center
($x$,$y$) and side length 9.  The infection pattern appears similar to
\FIGS\ref{fig:map}(a)-\ref{fig:map}(d)
 if we ignore the underlying space.  However, large
$L$, or the lack of global interactions, permits the disease to spread
only linearly in time \cite{Durrettbook}.  This contrasts with a
small-world type of networks and fully mixed networks like random
graphs in which diseases spread exponentially fast in the beginning
\cite{May_NAT_ROYAL,Newman_PRE99}. Accordingly, the transmission is by
far slower than shown in \FIGS\ref{fig:map}(a)-\ref{fig:map}(d). Although
propagations at linear rates would be good approximation before
long-range transportations had become readily available, they do not
match the recent spreads mediated by long-distance travelers that
lessen $L$ \cite{Riley,Abdullah,May_PRE,Watts_Wattsbook}. Taken in
another way, restrictions on long movements can be a useful spread
control \cite{Riley}.  By the same token, mathematical approaches such
as percolations and contact processes on regular lattices, which often
yield valuable rigorous results
\cite{Durrettbook,Schinazi_MB01,Liggettbook_Schinazibook}, are subject
to this caveat.

\subsection{Comparison with Scale-free Networks}

Another candidate for the network architecture is scale-free networks
whose distributions of $k_i$ obey the power laws
\cite{Barabasi_SCI}. Compared with the class of small-world networks
\cite{Watts_Wattsbook}, scale-free networks, particularly with the
original construction algorithm, lack the clustering property, whereas
they realize the power law often present in nature
\cite{Amaral_PNAS}. The chains of infection in a scale-free network
with the mean vertex degree equal to the previous simulations
are shown in \FIGS\ref{fig:map}(h) and \ref{fig:map}(i) for
$\overline{t}=1.0$ and $\overline{t}=2.0$, respectively.
Compared with the case of our transmission model
[see \FIGS\ref{fig:map}(a)-\ref{fig:map}(d)],
the influence of SS's is more magnified.
Figure~\ref{fig:trans}(b), plotting the distributions of $a_i$ for
$\overline{t}=2.0$,
shows that the distribution of $a_i$ decays with a power law rather than
exponentially for small $a_i$.  When more extensive data become
available, we will be able to fit \FIG\ref{fig:trans}(a) or
\ref{fig:trans}(b) to the real data as shown in
\FIG\ref{fig:trans}(c) and gain more insights into
the real epidemics, based on the
distributions of $a_i$. Figure~\ref{fig:trans} also
suggests that
more patients in total result from the epidemics 
in scale-free networks than in our model
network, even though the mean transmission rate and the mean vertex
degree are the same.

In \FIG\ref{fig:corr}, we plot
$(k_i,a_i)$ for each subpopulation of the susceptible ($a_i=0$), the
infected non-SS's, and the SS's.  For the infected non-SS's and SS's, $a_i$
is roughly proportional to $k_i$. This explains the power-law tail in
\FIG\ref{fig:trans}(b) and enables the existence of extremely
contagious SS's that could be called ultrasuperspreaders. The scale-free
property implies highly heterogeneous distribution of $k_i$.  Compared
with the same size of regular, small-world, or random networks whose
$k_i$'s are relatively homogeneous, scale-free networks have larger
$R_0\propto \left<k_i^2\right>/\left<k_i\right>$
\cite{May_NAT_ROYAL,Pastor,Lloyd_SCI,Newman_PRE02}.  In
percolation models, $R_0 = \sum^n_{i=0} k_i (k_i-1) \lambda_i$, where
$\lambda_i$ denotes the rate of possible transmission from the $i$th
individual \cite{Newman_PRE02}.  Consequently, in the original
scale-free networks whose density function of $k_i$ is proportional to
$k_i^{-3}$,
the critical value present for regular, small-world, or random
networks of the same mean edge density is extinguished
\cite{Newman_PRE02}. The same is true for dynamical models such as
contact processes \cite{Pastor}. Accordingly, scale-free networks
spread diseases even with infinitesimally small infection rates.
Furthermore, if a positive critical value exists with the type of
scale-free networks whose distribution of $k_i$ follows $k_i^{\gamma}$
($\gamma<-3$), a tendency that SS's occupy vertices with large $k_i$
can remove the critical values. For example, the critical infection
rate shrinks to 0 if $\lambda_i\propto k_i^{\gamma^{\prime}}$ with
$\gamma^{\prime} > -\gamma-3$.

Does this mechanism underlie
the current and possible spreading of SARS? We
think not, first because SS's do not necessarily 
seem to prefer to inhabit hubs of
networks.  Even without such correlation, heterogeneous infection
strengths of patients are not probably determined by the highly
heterogeneous $k_i$.  A major route for SARS transmission is
daily personal
contacts.  In this respect, distributions of $k_i$ of acquaintance
networks and friendship networks do not follow power laws, but have
exponential tails because of aging of individuals and their limited
capacity \cite{Amaral_PNAS,Lloyd_SCI}. Particularly, the number of
contacts per day is limited by the time and energy of a
person, which flattens the distribution of $k_i$; SS's 
of SARS seem to lead ordinary social lives. SS's possibly
result from the combination of large $\lambda_i$ and the
stay in groups with large $\left<k_i\right>_j T_j$, as has been discussed in this paper. Scale-free networks are rather relevant to spreads of
computer viruses and STD's 
\cite{Liljeros,May_PRE,Pastor,Lloyd_SCI}. Spreads are mostly 
mediated by individuals
on hubs in such epidemics, and ultrasuperspreaders may result as
a combination of large $\lambda_i$ and large $k_i$
\cite{May_NAT_ROYAL,May_PRE}. Preventive efforts to target
active patients with large $k_i$ are effective in these diseases
\cite{Newman_PRE02}. However, efforts to suppress SARS should be
invested in identifying the patients with large $\lambda_i$ and places
with large $\left<k_i\right>_j T_j$, rather than in looking for socially
active persons that exist only with probability exponentially small in
$k_i$.

\subsection{Effects of clustering}

A bonus of using a small-world type of networks
is that they are clustered, as measured by
the cluster coefficient $C$ \cite{Watts_Wattsbook}. In real
situations, the probability that two patients directly infected by the
same patient know each other is significantly high.  Also from this
viewpoint, small-world networks are more relevant than 
networks with small $C$ such as scale-free networks or random
graphs. We have used the network shown in
\FIG\ref{fig:network} instead of the model by Watts and Strogatz
\cite{Watts_Wattsbook} to facilitate analysis and 
comprehensive understanding of the dynamics.  With
edges appearing in different timings superimposed, $C \cong 
\left<k_i\right>/n_g c$
where $c$ is the number of random groupings ($c=2$ in our simulations),
whereas $L\propto \log n$. If $k_i$ is the order of $n_g$ and $c$ is not
so large, our network has small-world properties characterized
by large $C$ and small $L$.

The notion of clustering might induce one to imagine situations in
which people congregate and SARS spreads.  However,
infection occurs only on the boundaries between a susceptible
and a patient, and propagation slows if a pair of the infected
face each other as typically happens in highly clustered
networks. An increase in $C$ rather elevates the epidemic threshold in
site percolations \cite{Newman_PRE99,Moore}, bond percolations
\cite{Moore,Newman_PRE02}, and contact processes
\cite{Durrettbook,Keeling,Watts_Wattsbook}.  It also decreases the final
size of the infected population or spreads in late stages
\cite{Keeling,Watts_Wattsbook}. In spite of these general
effects of $C$, however, we
claim that $C$ does not count in the outbreak of SARS.
The possibility of outbreaks and dynamics in initial stages are determined
by other factors such as $\lambda_i$, $k_i$, $T_j$, and $\sigma$.  If
the $i$th individual that happens to be a patient has $\overline{k}$
neighboring patients, the effective $k_i$ decreases to
$k_i-\overline{k}$. However, $\overline{k}$ is tiny relative to $k_i$
in early stages even if $C$ is large.
On the other hand, clustering in the sense of large $C$ indirectly promotes
the spreads by increasing $k$. The arguments above on the effects of $C$ 
are based on varying $C$ with $k$ fixed.
However, the population density of a group
concurrently modulates $k$ and $C$ \cite{May_NAT_ROYAL}.
In a group of $n_g$ people with spatial size
$S_g$, $\left<k_i\right> =  (n_g-1)S_p/S_g$, where $S_p$ is
the size of personal space within which 
each person randomly interacts with others. Obviously, $\left<k_i\right>$
is proportional to the population density $n_g/S_g$. In addition,
$C = S_p/S_g\propto \left<k_i\right>$
even for a fully mixed population.
Therefore, the concept of clustering related to the SARS
spreads is high population density.
The network with large $C$ has been applied in this paper
to respect the social reality.

\section{Conclusions}
In this paper, we have proposed a dynamic network model for SARS
epidemics and shown that combined effects of superspreaders and
their possible tendencies to haunt potentially contagious places can
amplify the spreads. In addition, we have contrasted the different dynamical
consequences according to different types of underlying network structure.

\begin{acknowledgements}
We thank M. Urashima for helpful discussions.
This study is partially supported by the Japan Society for the
Promotion of Science and also by the Advanced and
Innovational Research Program in Life Sciences from the Ministry of
Education, Culture, Sports, Science, and Technology, the Japanese
Government.
\end{acknowledgements}


\bigskip
\bigskip

Figure captions

\bigskip

Figure 1: Schematic diagram of the dynamic network for $n_g=4$ and $g=4$.
The vertices initially form random graphs within each
group. After time $t_0$, they are randomly shuffled to reform new
groups. The graph switches between the two configurations
with period $t_0$.

\bigskip

\begin{figure}
\begin{center}
\includegraphics[height=1.25in,width=2.25in]{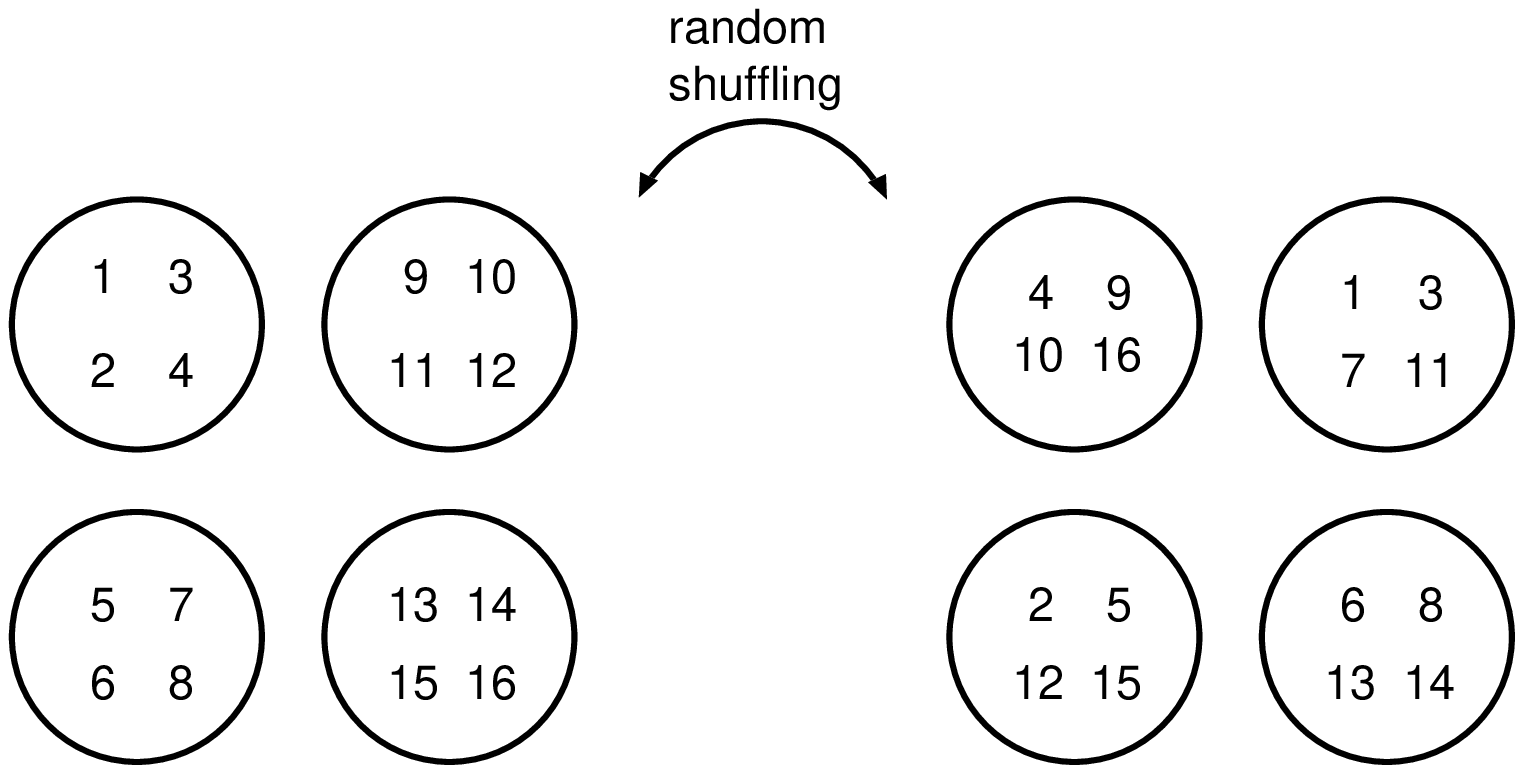}
\caption{}
\label{fig:network}
\end{center}
\end{figure}

\begin{figure}
\begin{center}
\includegraphics[height=1.8in,width=1.8in]{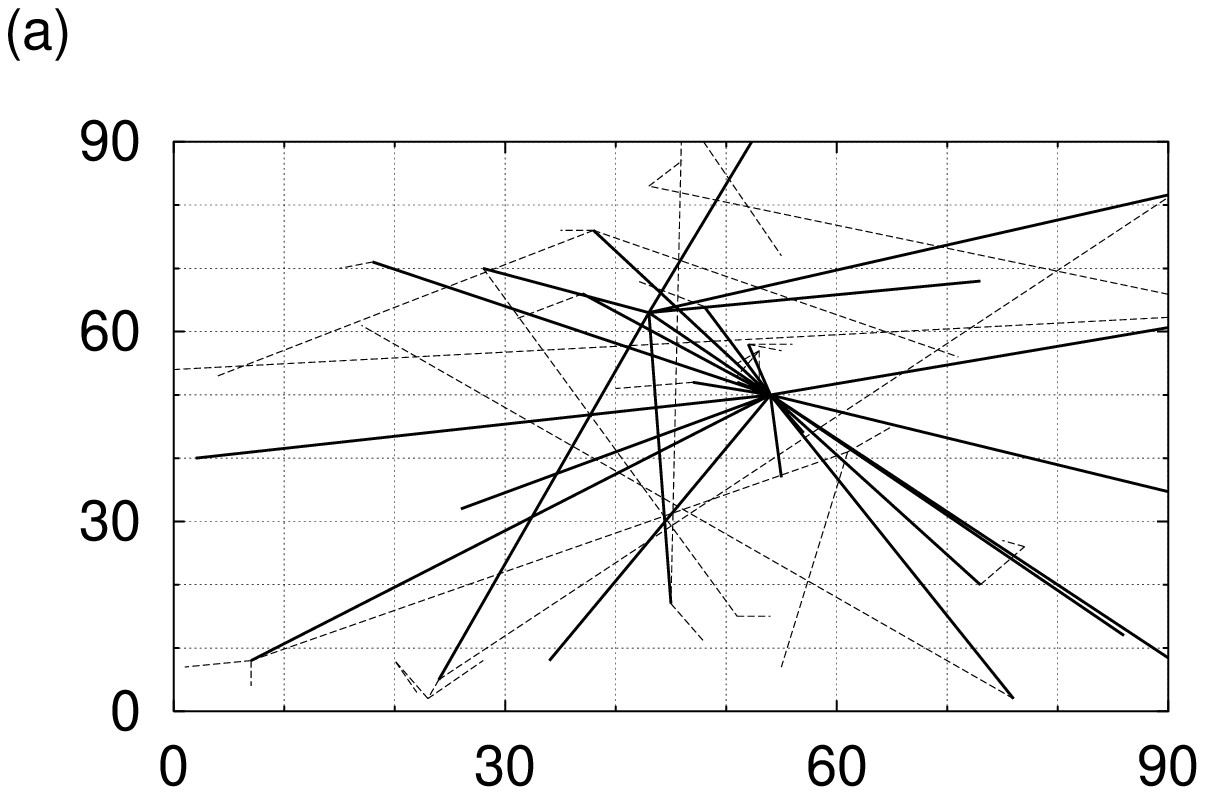}
\includegraphics[height=1.8in,width=1.8in]{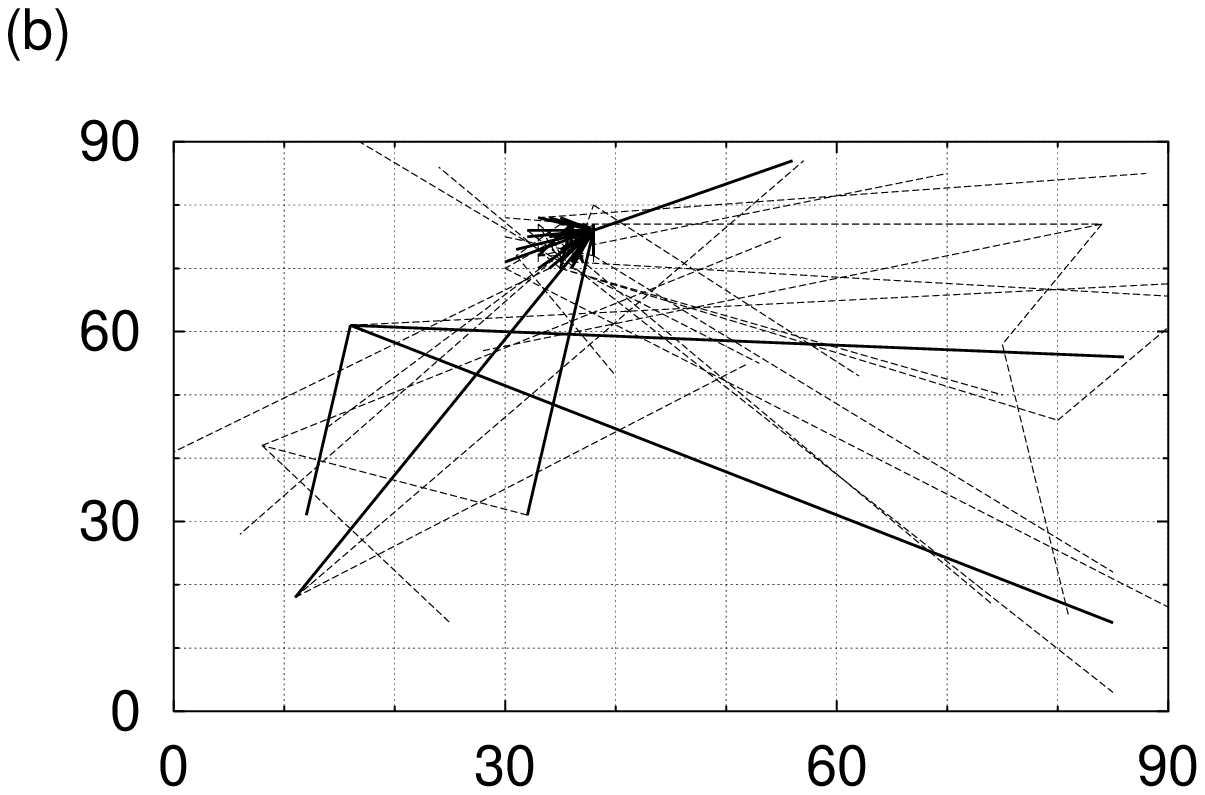}
\includegraphics[height=1.8in,width=1.8in]{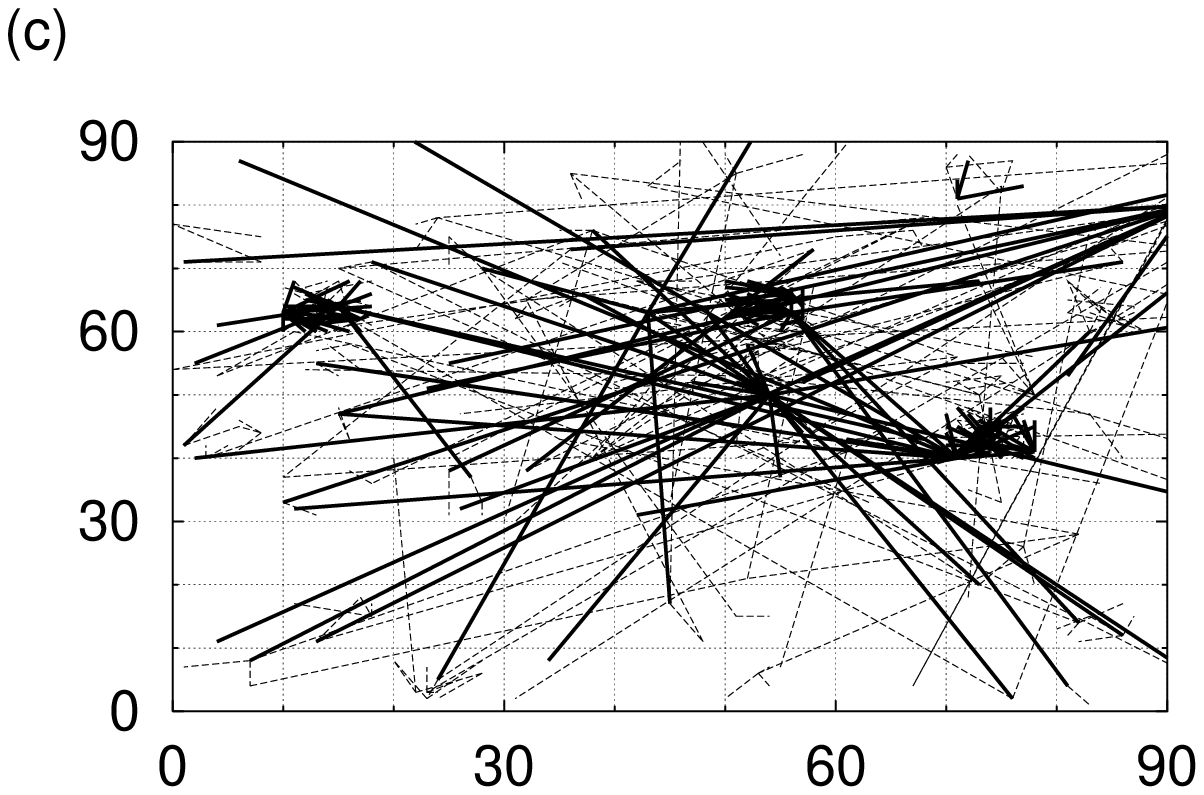}
\includegraphics[height=1.8in,width=1.8in]{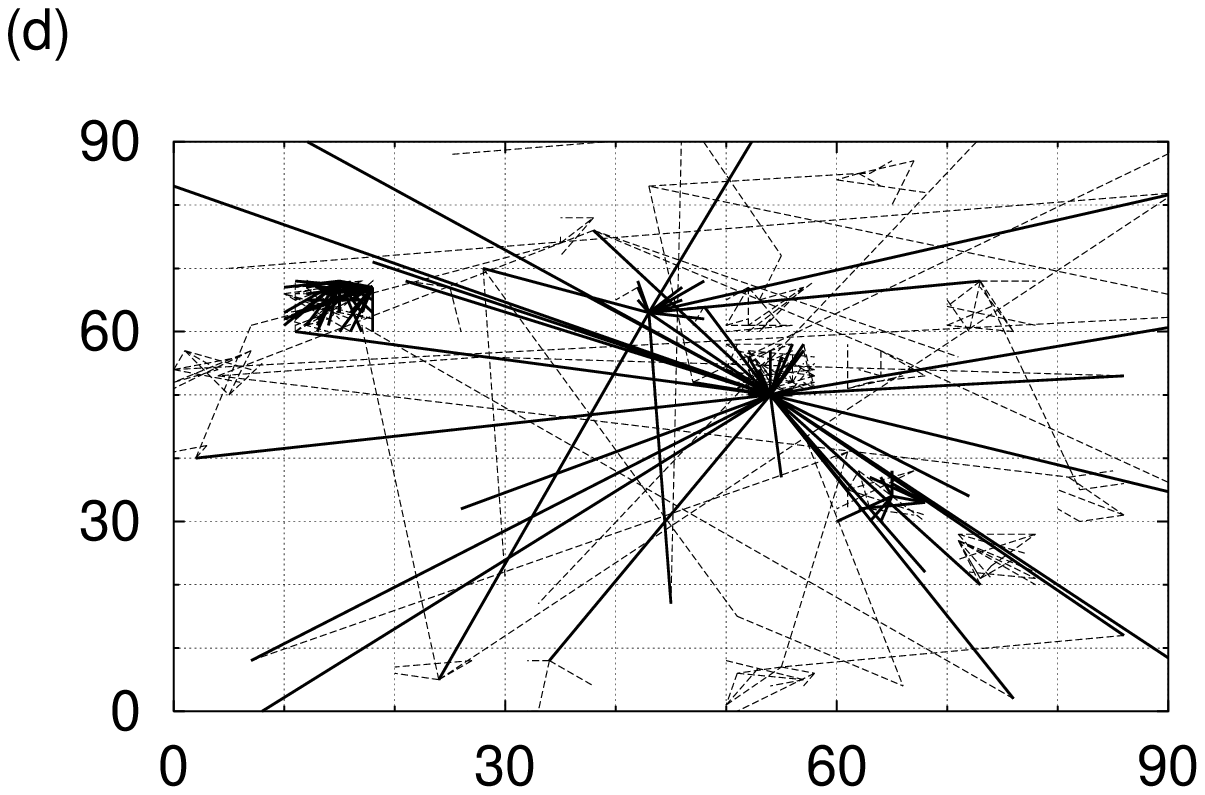}
\includegraphics[height=1.8in,width=1.8in]{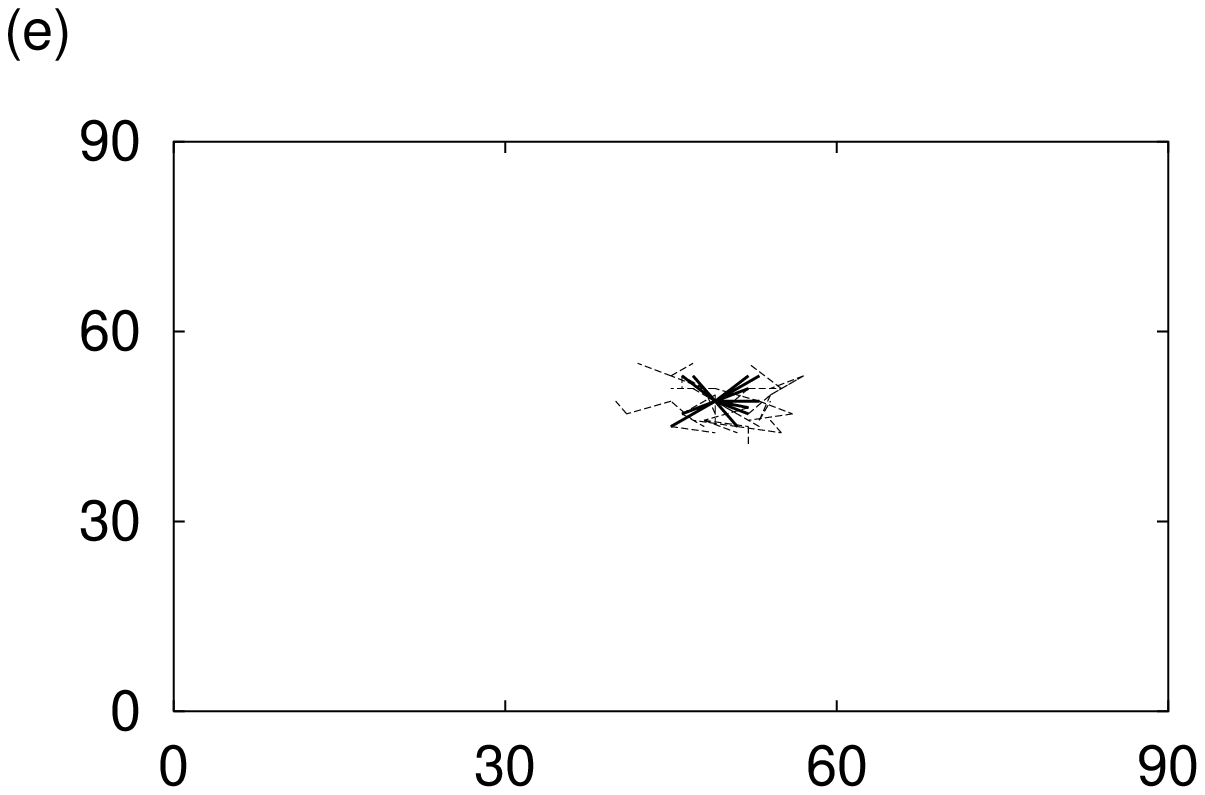}
\includegraphics[height=1.8in,width=1.8in]{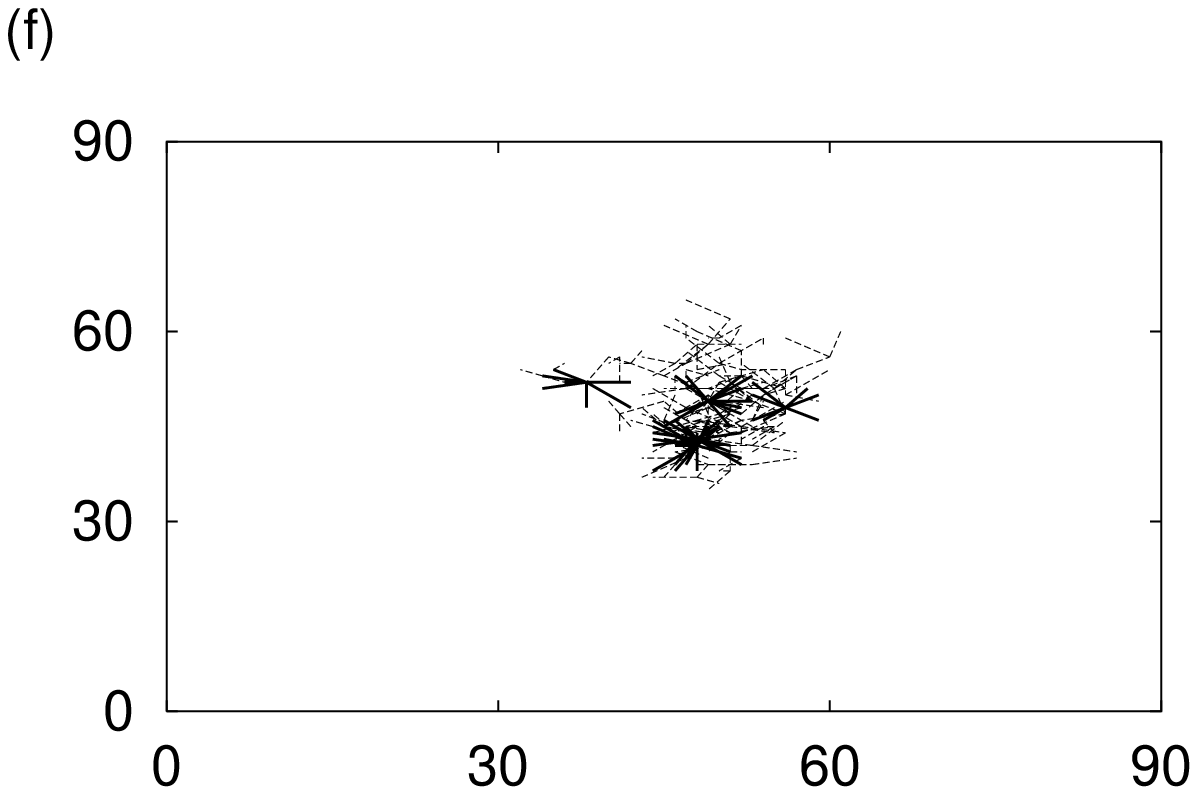}
\includegraphics[height=1.8in,width=1.8in]{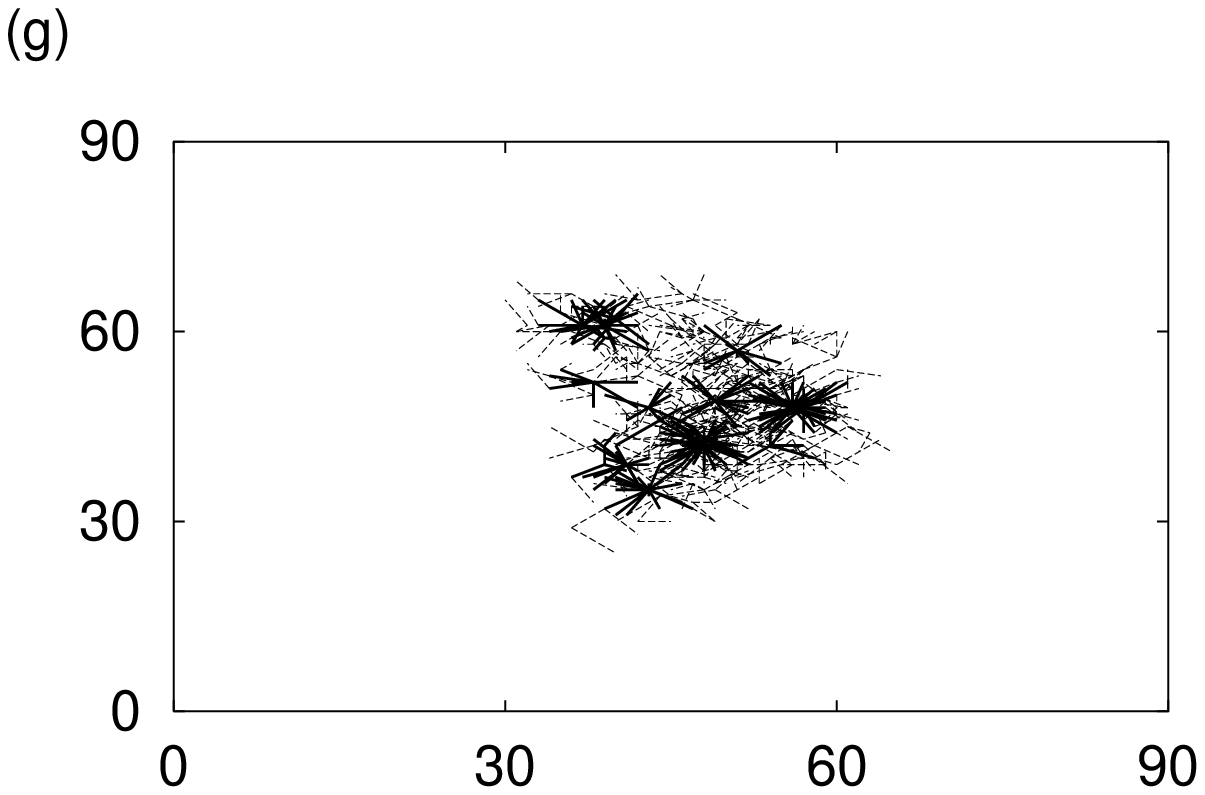}
\includegraphics[height=1.8in,width=1.8in]{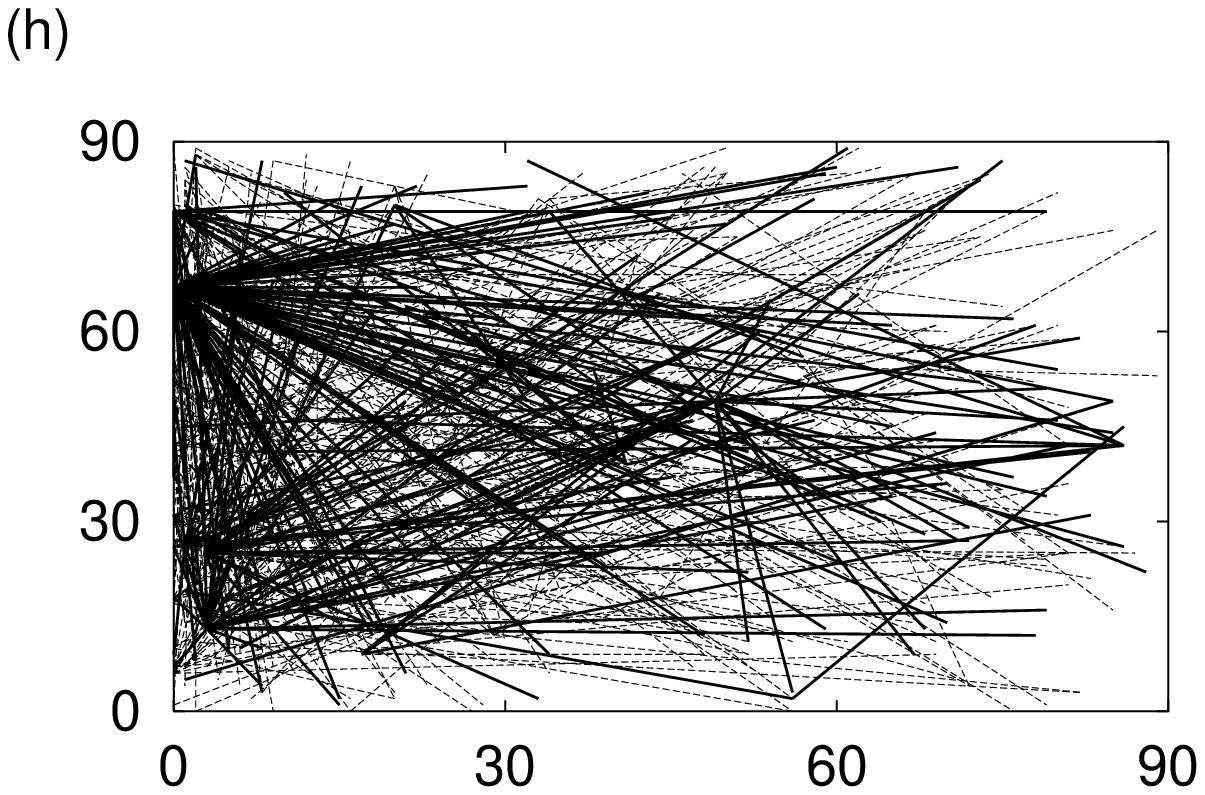}
\includegraphics[height=1.8in,width=1.8in]{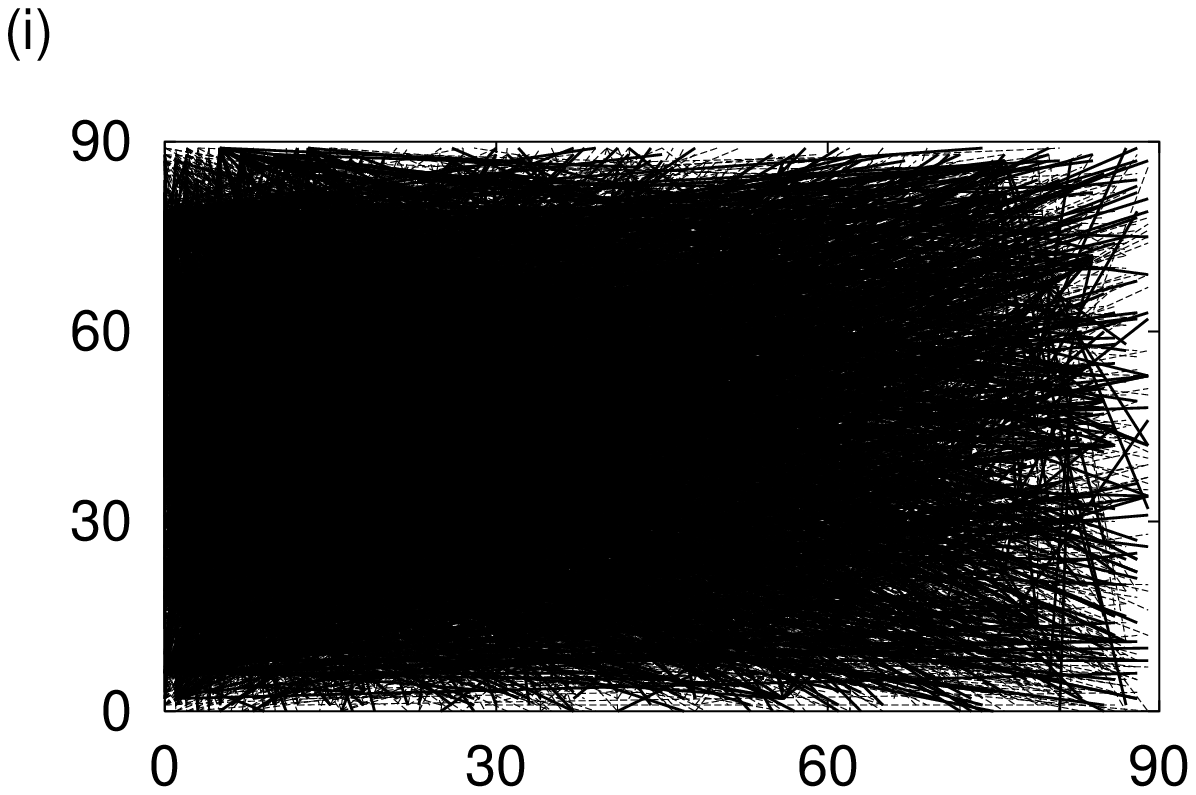}
\caption{}
\label{fig:map}
\end{center}
\end{figure}

\begin{figure}
\begin{center}
\includegraphics[height=2.25in,width=2.25in]{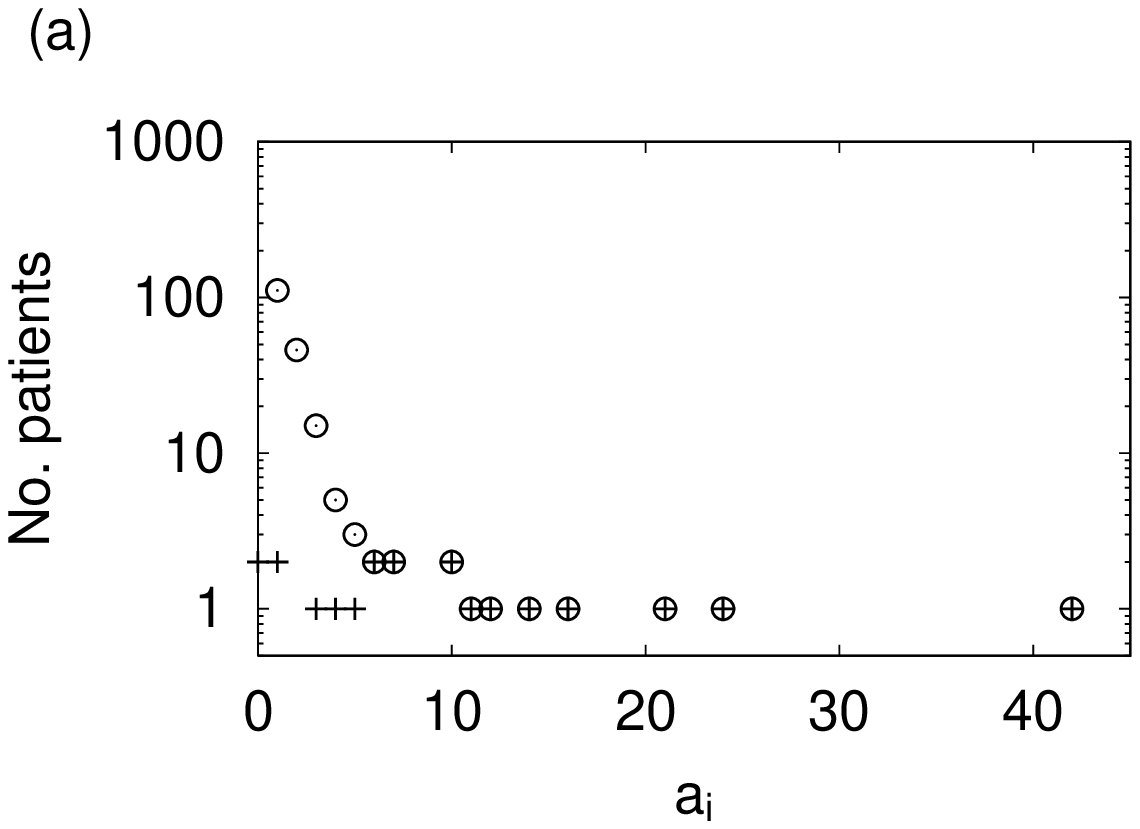}
\includegraphics[height=2.25in,width=2.25in]{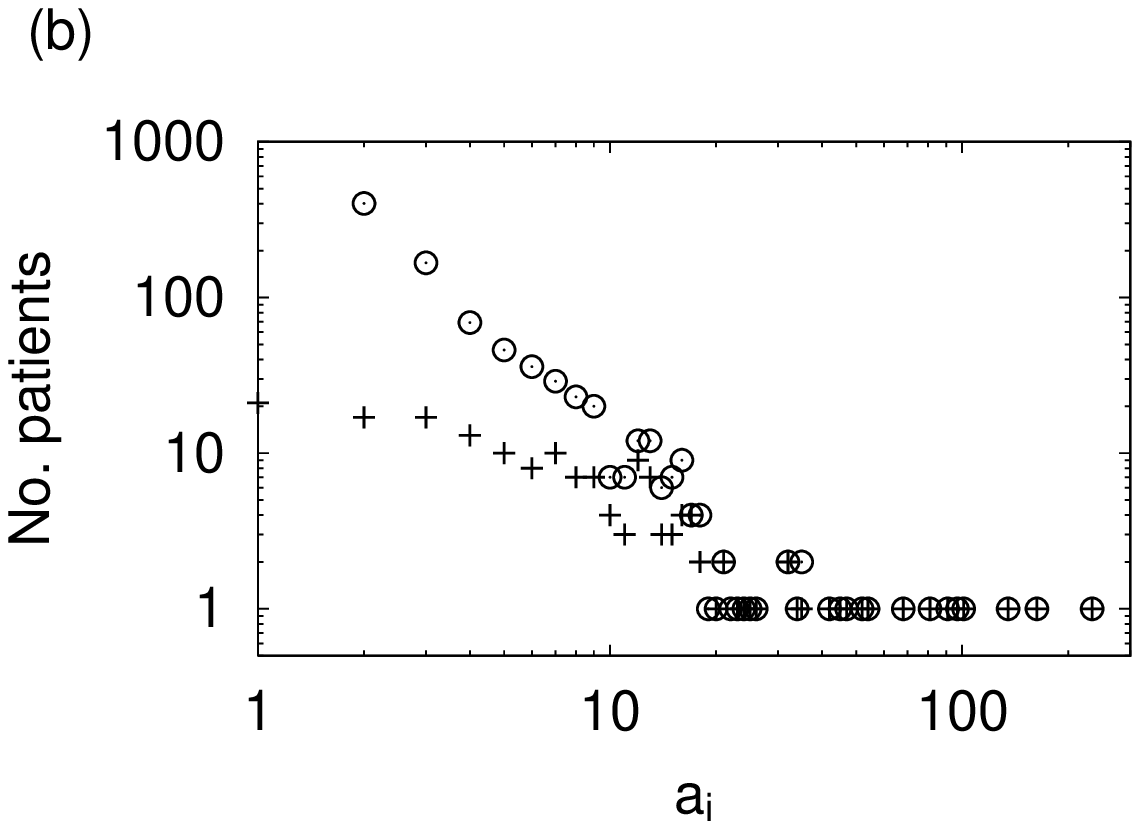}
\includegraphics[height=2.25in,width=2.25in]{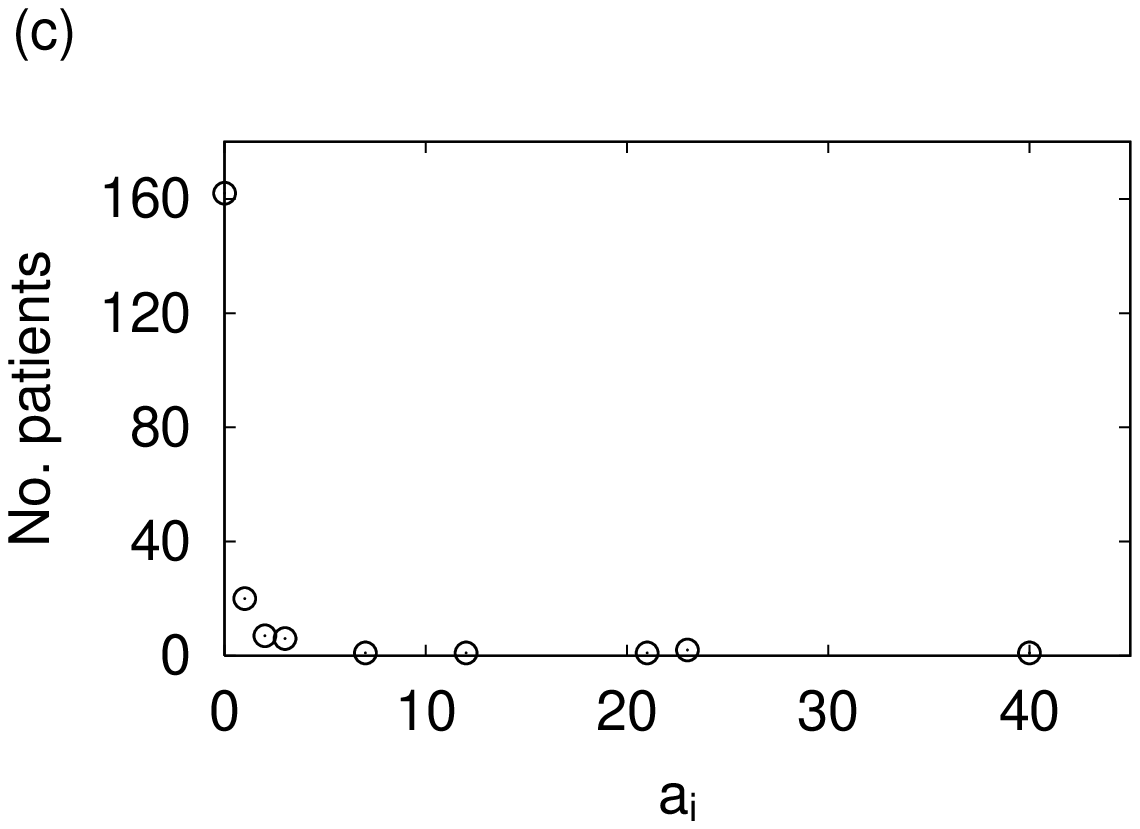}
\caption{}
\label{fig:trans}
\end{center}
\end{figure}

\begin{figure}
\begin{center}
\includegraphics[height=3.25in,width=3.25in]{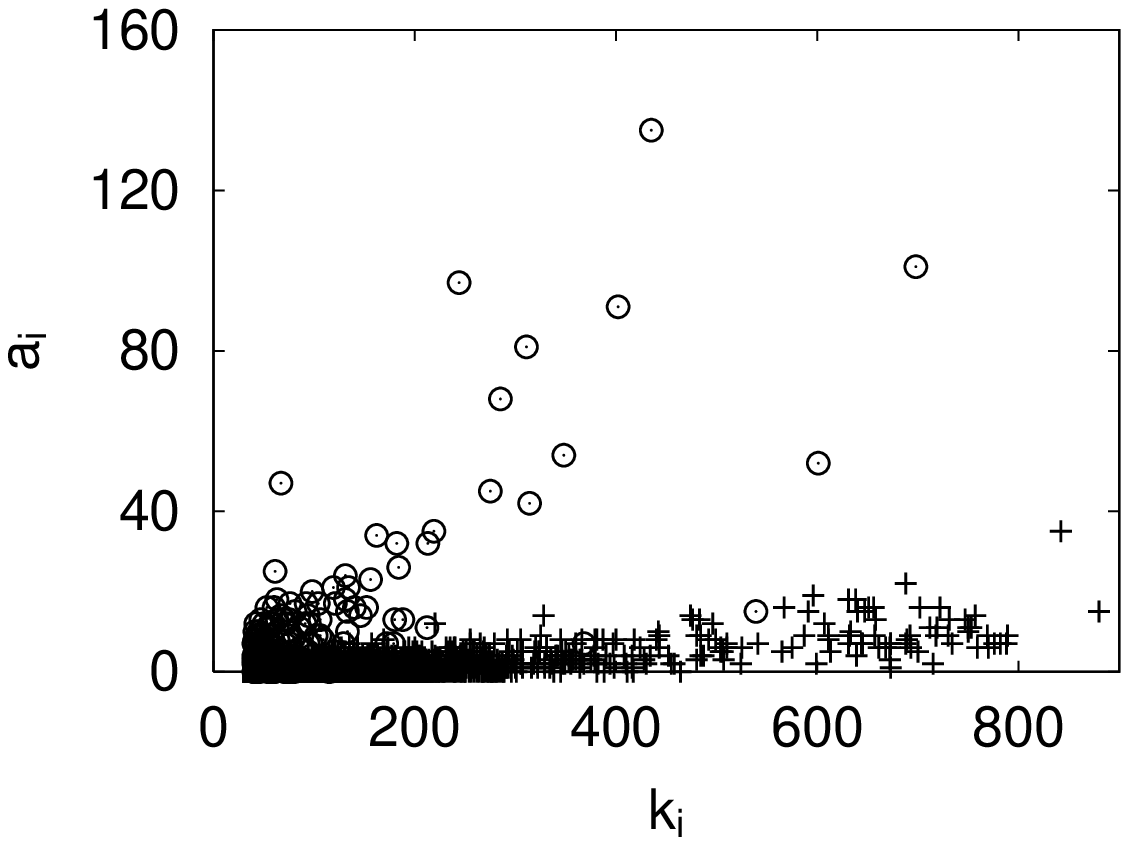}
\caption{}
\label{fig:corr}
\end{center}
\end{figure}

Figure 2: Chains of infection in the dynamical small-world network
(a), (b), (c), (d), the two-dimensional regular lattice (e), (f), (g),
and the scale-free network (h), (i). Transmissions from the infected
non-SS's and those from SS's are shown by dashed and solid lines,
respectively.  We set $n=90^2$, $n_g = 81$, $g=100$, $\lambda_I =
0.026$, $\lambda_{SS} = 0.52$, $k=80$, and the time step $\Delta t =
0.05$.  We set $t_0 = 0.5$ and $\overline{t}=1.0$ in (a), (b), $t_0=0.5$
and $\overline{t}=2.0$ in (c), $t_0=1.0$ and $\overline{t}=2.0$ in
(d), $\overline{t}=1.0$ in (e), (h), $\overline{t}=2.0$ in (f), (i), and
$\overline{t}=3.0$ in (g). (a) and (b) correspond to the two groupings
shown in \FIG\ref{fig:network}.  In (e), (f), (g), a square lattice with
$90\times 90$ vertices are used, and $k=80$.  In (h), (i), the
scale-free network with $k=80$ and $n=90^2$ is generated by starting
with a complete graph of 40 vertices and adding $n-40$ vertices. Each
vertex is endowed with 40 new edges whose destinations are determined
according to preferential attachment \cite{Barabasi_SCI}.

\bigskip

Figure 3: Distributions of $a_i$ --- namely, the number of individuals
to whom a patient has directly infected --- in (a) the dynamical small-world
network, (b) the scale-free network, and (c) Singapore
\cite{Singapore}. The distributions are shown for the SS's
(crosses) and all the patients (circles).  We set $\overline{t}=2.0$
in (a), (b) and $t_0=0.5$ in (a).

\bigskip

Figure 4: Relation between the vertex degree $k_i$ and the number of
infections, $a_i$, in the scale-free network for the susceptible
(squares), the infected non-SS's (crosses), and the SS'sars.tex: main
text (PRE)

\clearpage

\end{document}